\documentclass{article}
\usepackage[preprint]{neurips_2025}

\usepackage[utf8]{inputenc}
\usepackage[T1]{fontenc}
\usepackage[hypertexnames=false,colorlinks=true,linkcolor=blue,citecolor=blue,urlcolor=blue]{hyperref}
\usepackage{url}
\usepackage{booktabs}
\usepackage{amsfonts}
\usepackage{amsmath}
\usepackage{amssymb}
\usepackage{microtype}
\usepackage{xcolor}
\usepackage{graphicx}
\usepackage{multirow}
\usepackage{array}
\usepackage{tabularx}
\usepackage{longtable}
\usepackage{float}

\newcommand{\sys}{MISA-T}
\newcommand{\vllmroutercell}{vLLM Router}
\newcommand{\vllmrouterhighcell}{\mbox{vLLM Router (high load)}}

\definecolor{abstractbg}{RGB}{236,247,255}
\newcommand{\tablefont}{\footnotesize}
\newcolumntype{Y}{>{\raggedright\arraybackslash}X}

\title{Scheduling Mixed RL Rollouts Beyond Prefix Locality}

\author{%
  \normalfont
  Zetao Hong\textsuperscript{1} \quad
  Song Yuan\textsuperscript{2} \quad
  Yuanhao Ding\textsuperscript{2} \\
  Yibo Zhu\textsuperscript{2} \quad
  Daxin Jiang\textsuperscript{2} \quad
  Zhibin Wang\textsuperscript{1,*} \quad
  Chen Tian\textsuperscript{1} \\
  \textsuperscript{1}State Key Laboratory of Novel Software Technology, Nanjing University \\
  \textsuperscript{2}StepFun
}

\begin{document}

\maketitle
\begingroup
\renewcommand{\thefootnote}{\fnsymbol{footnote}}
\footnotetext[1]{Corresponding to wzbwangzhibin@gmail.com}
\endgroup

\begin{center}
\begingroup
\setlength{\fboxsep}{10pt}
\colorbox{abstractbg}{%
\begin{minipage}{0.90\linewidth}
Modern reinforcement learning (RL) post-training pipelines for large language models (LLMs) increasingly combine rollout workloads across multiple domains and feedback paradigms.
Prefix-aware routing improves inference efficiency through cache reuse and load balancing, but it does not control how heterogeneous rollout sessions compete for KV-cache capacity.
When reinforcement learning with verifiable rewards (RLVR), reinforcement learning from human feedback (RLHF), and agentic rollouts share an asynchronous inference service, their distinct sequence structures, interaction patterns, and KV-residency times create substantially different serving demands.
Rollout scheduling must account for this heterogeneity without distorting the workload mixture specified by the trainer.

We present \textbf{\sys{}}, a routing-layer admission policy for mixed rollout serving.
\sys{} combines adaptive session admission, workload-aware KV-capacity allocation, and residency-time-aware KV accounting.
In rollout-only ablations on Step3.7 and Qwen3.6-35B-A3B, \sys{} improves rollout throughput over a sweep-tuned cache-aware vLLM Router by $53.3\%$ and $43.6\%$, respectively, while maintaining high prefix-cache hit rates.
In a matched 50-iteration Step3.7 experiment, it increases rollout throughput by $35.6\%$ and reduces mean iteration time by $22.8\%$, while keeping the consumed workload mixture close to the trainer target and achieving comparable task scores.
\end{minipage}}
\endgroup
\end{center}

\section{Introduction}

Modern RL post-training for large language models increasingly treats rollout generation as a distributed serving workload.
A trainer releases prompts or agent tasks, the rollout layer translates them into model-generation requests, and a router assigns those requests to inference instances that reuse cached KV state when possible.
Existing LLM serving systems therefore treat prefix reuse as a central efficiency objective through PagedAttention-based KV memory management~\citep{kwon2023vllm}, RadixAttention-based reuse~\citep{zheng2024sglang}, and distributed prefix-aware routing~\citep{srivatsa2025preble,cao2025localityaware,dexter2025prefixreuse}.
Reusing cached prefixes reduces redundant prefill whenever requests share prompt or session context.

The scheduling problem becomes more complex when a rollout stream contains heterogeneous workloads.
Such streams arise in multi-domain RL~\citep{yang2025qwen3,liang2025modomodo} and in multi-teacher on-policy distillation stages that integrate domain specialists~\citep{ma2026mopd,yang2026exopd,li2026katcoderv2,kimiteam2026k3,yang2026nemotroncascade2}.
When served by a common inference pool, these workloads stress different resources.
Rollout pipelines commonly fan out multiple samples per prompt, creating group-completion dependencies across workload classes~\citep{yu2026multirollout}.
Within this shared structure, RLVR requests often have short prompts and long decode tails; RLHF requests have more balanced prompt and response lengths; and agentic trajectories depend on session continuity across repeated long-prefix calls and intervening tool execution.
For agentic rollouts in particular, a cache miss on a later turn requires prefill of the accumulated interaction history~\citep{kang2026thunderagent,zhu2026tracelab}.

The trainer releases rollout tasks subject to staleness and maximum-concurrency limits, whereas the router controls serving-side admission and instance selection.
Prefix reuse remains important, but selecting the instance with the largest cached prefix alone does not control how many sessions compete for limited KV-cache capacity or how that capacity should be shared across workloads.
The router must therefore account for workload-specific serving characteristics without altering the workload composition defined by the trainer.

We formulate this setting as mixed-rollout routing under trainer-side mixture constraints and propose \sys{} (\textbf{M}ix-aware \textbf{S}ession \textbf{A}dmission with a \textbf{T}ime factor).
\sys{} separates the trainer's mixture contract from the router's resource controller: the trainer determines the target distribution, while the router performs session-aware admission control, partitions protected KV capacity by workload class, and accounts for class-specific KV residency time.
The contributions are:
\begin{enumerate}
    \item \textbf{Overload-aware session admission.} We derive an adaptive admission cap for new sessions from observed KV demand and overload pressure. Requests beyond the cap are held and periodically re-evaluated, while existing continuations remain protected, limiting KV-cache churn and reducing the need to retune static concurrency limits.
    \item \textbf{Workload-aware capacity allocation.} We partition protected admission capacity across RLVR, RLHF, and agentic workloads according to their demand and KV footprints. This allows resource control to reflect heterogeneous sequence structures while leaving the target workload composition under trainer control.
    \item \textbf{Residency-time-aware KV accounting.} We weight each workload's KV demand by its observed session residency time, capturing both inference time and tool-interaction intervals during which reusable KV remains resident. This extends footprint-based allocation to account for KV block-time demand.
\end{enumerate}
Across two rollout-only model families, \sys{} improves rollout throughput by up to $53.3\%$ over a sweep-tuned vLLM Router baseline.
In matched end-to-end training, it improves rollout throughput by $35.6\%$ and reduces mean iteration time by $22.8\%$.

\section{Background and Motivation}

\subsection{Rollout serving architecture}

\begin{figure}[!t]
    \centering
    \includegraphics[width=0.98\linewidth]{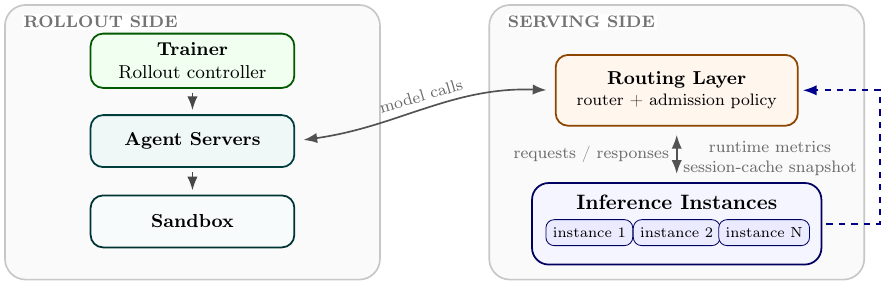}
	    \caption{System overview. \sys{} is a routing-layer policy. It can be deployed with the router and scheduler as separate services or as a gateway plugin, and depends only on request metadata, runtime metrics, and session-cache snapshots from the serving layer.}
	    \label{fig:system}
\end{figure}

Figure~\ref{fig:system} illustrates a common architecture for distributed RL rollout generation.
A trainer releases rollout tasks and bounds in-flight work through maximum-concurrency and policy-version staleness limits.
In fully asynchronous systems, rollout generation proceeds concurrently with model training~\citep{fu2025areal,yan2025arealhex,hu2026dora}.
Rollout workers translate these tasks into model requests, while a routing layer performs request admission and selects an inference instance.
For agentic workloads, an agent server additionally maintains the interaction loop and executes tool actions in an isolated sandbox.

This paper focuses exclusively on the routing layer.
Although the router and request scheduler are separate services in our deployment, they form one logical component and can also be implemented as a gateway plugin.
\sys{} relies on request metadata, runtime metrics, and lightweight session-cache snapshots exposed by the serving layer.
It therefore composes with existing rollout controllers and inference engines while leaving model execution, physical KV-cache allocation, and the agent runtime unchanged.

\subsection{Mixed rollout generation}

We study mixed-workload rollout generation, rather than a specific RL algorithm.
A mixed rollout stream can arise from different training organizations.
Joint multi-domain RL samples heterogeneous prompts within one on-policy process~\citep{yang2025qwen3,liang2025modomodo}.
Staged pipelines train domain specialists separately, but their integration stage can again mix domains: MOPD samples prompts from a multi-domain dataset, generates student rollouts, and routes each trajectory to its matching teacher~\citep{ma2026mopd}.
This capability-integration pattern is used in MiMo-V2-Flash~\citep{xiao2026mimov2flash}, Kimi K3 across domain and reasoning-effort specialists~\citep{kimiteam2026k3}, Nemotron 3 Ultra~\citep{nvidia2026nemotronultra}, and Nemotron-Cascade 2~\citep{yang2026nemotroncascade2}.
These deployments indicate that mixed-domain on-policy rollout generation is becoming a recurring post-training workload.
These training organizations do not prescribe how rollout generation is executed.
Across these settings, the trainer defines the sampling distribution and rollout-group semantics.
The serving layer observes cache state and queueing pressure that change at request time scale.
Its policy should improve serving efficiency without redefining the workload composition specified by the trainer.

\subsection{Prefix-aware routing}

Prefix-cache-aware routing is well established in LLM serving.
vLLM improves KV memory management with PagedAttention~\citep{kwon2023vllm}; SGLang introduces RadixAttention for reusable KV prefixes across structured generation calls~\citep{zheng2024sglang}; Preble balances distributed prefix reuse against load~\citep{srivatsa2025preble}; DLPM/D$^2$LPM combines prefix reuse with fairness constraints~\citep{cao2025localityaware}; and k-LPM incorporates latency constraints~\citep{dexter2025prefixreuse}.
Collectively, these systems establish prefix-aware load balancing as a common routing design for distributed rollout serving.

The cache-aware policy in vLLM Router belongs to this family~\citep{vllmrouter2026}.
vLLM Router controls placement but not admission: it chooses an inference instance for each request but does not limit how many new sessions enter that instance.
Under high concurrency, these sessions can compete for KV capacity and evict reusable prefixes.
More broadly, the routing systems discussed above primarily optimize placement, load balancing, or prefix reuse; they do not allocate admission capacity according to workload-class KV block-time demand.
This gap becomes important when heterogeneous rollout workloads share an inference pool.
Prefix-cache hit rate remains a useful indicator of avoided prefill work, but it cannot by itself characterize rollout-serving efficiency.
We therefore evaluate it together with request throughput, iteration time, and consumed workload mixture.

\subsection{Workload asymmetry}
\label{sec:workload-characterization}

Mixed rollout streams can place workloads with substantially different serving behavior on the same inference pool.
Table~\ref{tab:workload-diff} summarizes the profiles observed in our trace.
These profiles characterize this deployment rather than define RLVR, RLHF, or agentic training: turn structure and token lengths vary with the model, dataset, and environment.

\begin{table}[!htbp]
    \centering
    \caption{Serving characteristics of mixed rollout workloads.}
    \label{tab:workload-diff}
    \tablefont
	    \begin{tabularx}{\linewidth}{lYYY}
        \toprule
        Characteristic & RLVR & RLHF & Agentic \\
        \midrule
        Observed turns & Single-turn & Single-turn & Multi-turn \\
        Token lengths & Short input; long output & Medium input and output & Long input; short output \\
        KV behavior & Decode-time growth & Moderate residency & Retained between turns \\
        Dominant cost & Long decode & Prefill and decode & Prefix reuse; residency \\
        \bottomrule
    \end{tabularx}
\end{table}

In our trace, RLVR is decode-dominated, RLHF has intermediate prompt and response lengths, and agentic rollouts retain long accumulated prefixes across tool-interleaved turns.
These patterns are consistent with prior studies of long-output reasoning, conventional RLHF, and agentic serving~\citep{du2025ulorl,ouyang2022instructgpt,kang2026thunderagent,zhu2026tracelab}.

Table~\ref{tab:prelim-stats} quantifies these differences in the online mixed-rollout trace with a 128K-token context limit.

\begin{table}[H]
    \centering
    \caption{Request-level token statistics under a 128K context limit.}
    \label{tab:prelim-stats}
    \tablefont
    \setlength{\tabcolsep}{2.8pt}
    \begin{tabular*}{\linewidth}{@{\extracolsep{\fill}}lrrrrrrrr}
        \toprule
        Workload & Requests & Sessions & Input p50 & Input p95 & Input p99 & Output p50 & Output p95 & Output p99 \\
        \midrule
        Agent & 73,659 & 1,058 & 37,757 & 103,209 & 123,207 & 148 & 1,612 & 3,930 \\
        RLHF & 599 & 599 & 60 & 2,564 & 2,564 & 7,344 & 24,447 & 37,101 \\
        RLVR & 809 & 809 & 156 & 1,129 & 1,277 & 47,583 & 130,815 & 130,958 \\
        \bottomrule
    \end{tabular*}
\end{table}

The trace therefore contrasts residency-heavy agent sessions, which reuse long prefixes across tool-interleaved turns, with decode-heavy RLVR samples dominated by long outputs.

Table~\ref{tab:agent-stage-breakdown} further separates model execution from tool interaction in the agent workloads.
The larger tool gap for Qwen3.6-35B-A3B shows why model inference time alone can underestimate KV-cache residency.
Here, tool gap denotes tool-use time outside model inference.

\begin{table}[H]
    \centering
    \caption{Agent task-state occupancy.}
    \label{tab:agent-stage-breakdown}
    \tablefont
    \setlength{\tabcolsep}{3.6pt}
    \begin{tabular*}{\linewidth}{@{\extracolsep{\fill}}lrr}
        \toprule
        Model & Model request & Tool/orchestration gap \\
        \midrule
        Step3.7 & 93.4\% & 6.6\% \\
        Qwen3.6-35B-A3B & 77.3\% & 22.7\% \\
        \bottomrule
    \end{tabular*}
\end{table}

\subsection{Why placement-only routing falls short}

Our measurements expose a failure mode that placement alone cannot prevent.
Placement-only routing can collapse under a concurrency ceiling that admission-controlled policies tolerate.
On Qwen3.6-35B-A3B, raising vLLM Router to this high-load ceiling reduces request throughput by $52.5\%$, effective prefill throughput by $57.3\%$, and decode throughput by $38.5\%$ relative to its sweep-tuned operating point; prefix-cache hit rate falls from $92.4\%$ to $4.5\%$.
The failure forms a positive feedback loop: admitting more new sessions expands the resident KV working set, evicts reusable continuations, converts later turns into cold prefill, and lengthens the queue and session overlap, which creates further eviction pressure.

A global session cap can interrupt this loop, but it still treats heterogeneous sessions as interchangeable.
The workload measurements in Section~\ref{sec:workload-characterization} show why aggregate overload control is insufficient: classes differ substantially in both KV footprint and residency time, so a shared cap does not determine how protected capacity should be divided among them.
Under the same mixed stream and permissive concurrency ceiling, \sys{} improves rollout throughput over class-agnostic Session Admission by $37.3\%$ on Step3.7 and $20.4\%$ on Qwen3.6-35B-A3B.
We separate the contributions of workload-specific allocation and residency-time weighting in Section~\ref{sec:evaluation}.

\subsection{Key insight and system challenges}

\paragraph{Admission is a KV commitment.}
Admitting a new session is not an isolated request-placement decision; it commits capacity for subsequent KV growth, future continuations, and intervals between turns.
At a scheduling epoch, a backlog of $N_b$ unfinished sessions with mean KV footprint $\bar{k}_b$ and mean residency time $T_b$ represents approximately $N_b\bar{k}_bT_b$ block-time work.
A per-class block quota bounds how much protected KV each class can commit, and dividing that quota by its footprint yields an enforceable session cap.
The cap does not prescribe trainer sampling or guarantee a fixed class throughput; it prevents one class from exhausting the protected working set needed by others.

This observation leads to three systems challenges:
\begin{enumerate}
    \item \textbf{Online admission without repeated concurrency tuning.} A static operating point can become inaccurate as the model, sequence lengths, or workload mix changes. Because final sequence growth and residency are unknown and cache observations arrive asynchronously, the scheduler must regulate admission from delayed state while reducing reliance on configuration-specific concurrency sweeps.
    \item \textbf{Accounting for workload heterogeneity.} Classes differ in KV footprint and residency, yet the serving policy must not change the workload composition specified by the trainer.
    \item \textbf{Composition with cache-aware serving.} Admission must coexist with cache-aware placement and optional KV offloading without taking over their placement or cache-management responsibilities.
\end{enumerate}

\section{Problem Formulation}

The measurements above separate two routing decisions.
Placement chooses where an admitted request can reuse prefix state, whereas admission bounds the future population of cache-resident sessions.
We now formalize these decisions under the trainer's workload contract and a protected KV-capacity constraint.

\subsection{Trainer contract and routing decisions}

We consider a trainer that releases rollout units from workload classes $\mathcal{B}$ to a shared pool of inference instances $\mathcal{W}$.
A rollout unit is one trainer-counted sample, and a prompt may fan out into several grouped units.
For agentic workloads, each unit is a multi-turn trajectory with interleaved model requests and tool execution.
Each request $r$ inherits a workload class $b_r\in\mathcal{B}$, rollout-unit identity $u_r$, session identity $s_r$, and optional group identity $g_r$ from the trainer.
The trainer controls policy staleness, maximum concurrency, and target mixture $\rho\in\Delta(\mathcal{B})$; the router neither creates rollout units nor changes their labels.
Let $A_b(T)$ denote the class-$b$ rollout units released by time $T$ and $A(T)=\sum_b A_b(T)$, and assume their empirical proportions converge to the trainer-specified mixture $\rho$.
For each request, it chooses $\pi_t(r)\in\mathcal{W}\cup\{\mathrm{HOLD}\}$: selecting $w\in\mathcal{W}$ admits and places the request on instance $w$, whereas \textsc{Hold} delays admission and periodically re-evaluates the request.
A continuation reuses retained KV, whereas admitting a new session commits protected capacity for its subsequent growth and future continuations.
These decisions are online.
At admission time, the router observes request metadata and current serving state but not final sequence length, future turns, or session residency.
The policy must therefore make admission and placement decisions from prefix state, runtime metrics, session-cache snapshots, and bounded histories.

\subsection{Prefix reuse and KV block-time demand}

For request $r$, let $L_r$ be its prompt length and $H_{r,w}(t)$ the reusable prefix on inference instance $w$.
We model the estimated prefill cost on instance $w$ as
\begin{equation}
\widehat{c}_{r,w}(t)=d_w(t)+\frac{L_r-H_{r,w}(t)}{\mu_w(t)}.
\label{eq:prefill-cost}
\end{equation}
Here, $d_w(t)$ is the estimated prefill backlog delay in seconds and $\mu_w(t)$ is the measured prefill throughput.
This placement rule balances prefix reuse against load, while admission controls how many sessions enter the cache.

For session $s$, let $k_s(t)$ be its retained KV blocks and $T_s$ its effective residency duration.
Its KV occupation is naturally measured in block-time,
\begin{equation}
Z_s=\int k_s(t)\,dt\ \approx\ \bar{k}_s T_s,
\label{eq:session-block-time}
\end{equation}
where $\bar{k}_s$ is a conservative reference footprint.
This approximation captures both retained cache size and useful residency; for agentic sessions, residency includes tool-execution intervals between turns.

If $\mathcal{P}_w(t)$ is the set of sessions protected on instance $w$, $\widetilde{k}_s(t)$ is the controller's reserved-footprint estimate, and $C_w$ is the protected budget, admissible state satisfies
\begin{equation}
\sum_{s\in\mathcal{P}_w(t)}\widetilde{k}_s(t)\leq C_w,
\qquad \forall w\in\mathcal{W}.
\label{eq:protected-kv-constraint}
\end{equation}

\subsection{Scheduling objective under the trainer contract}

Let $N_b^\pi(T)$ be the number of complete class-$b$ rollout samples delivered to the trainer by time $T$ under policy $\pi$.
We define rollout throughput as
\begin{equation}
S_\pi(T)=\frac{1}{T}\sum_{b\in\mathcal{B}}N_b^\pi(T).
\label{eq:rollout-completion-rate}
\end{equation}
An agentic trajectory counts once after all turns and tool interactions complete, so $S_\pi(T)$ differs from request throughput.
Define the class backlog as $Q_b^\pi(T)=A_b(T)-N_b^\pi(T)$.
For a feasible offered load, we use the following design objective:
\begin{equation}
\max_{\pi}\ \liminf_{T\rightarrow\infty}S_\pi(T)
\quad\text{subject to}\quad
Q_b^\pi(T)=o\!\left(A(T)\right),\qquad \forall b\in\mathcal{B}.
\label{eq:scheduling-objective}
\end{equation}
The protected-capacity constraint in Eq.~\eqref{eq:protected-kv-constraint} also applies.
The backlog condition asks the routing layer not to persistently delay any offered class.
Because $N_b^\pi=A_b-Q_b^\pi$, if $A_b/A\rightarrow\rho_b$ and $Q_b^\pi/A\rightarrow0$, then the completed mixture also converges to $\rho$.
MISA-T approximates this objective online through backlog-dependent quotas; for finite runs, we report $D_{\mathrm{TV}}(p,q)=\frac{1}{2}\lVert p-q\rVert_1$ between the completed and target mixtures.

\section{Mix-Aware Session Admission}

\begin{figure}[!t]
    \centering
    \includegraphics[width=\linewidth]{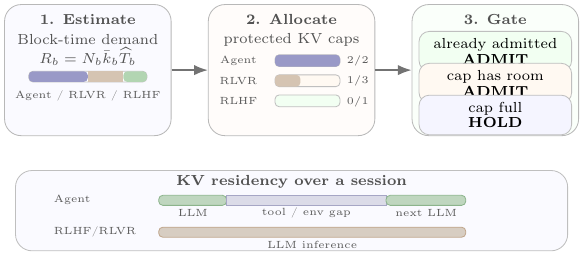}
\caption{\sys{} scheduling overview.
Existing admitted sessions preserve locality, while each additional admitted session commits capacity for future KV growth and reuse.}
    \label{fig:principle}
\end{figure}

Building on the trainer contract in Section~3, \sys{} combines adaptive session admission, workload-aware KV-capacity allocation, and residency-time-aware accounting.
The controller changes only request-admission timing and inference-instance placement.
Figure~\ref{fig:principle} summarizes this admission flow and the session-residency signal used by the controller.

\subsection{Adaptive session admission}

vLLM Router determines where a request should run, but not how many distinct sessions may occupy an inference instance's protected KV-cache capacity.
For inference instance $w$, let $K_w(t)$ denote the protected session cap, $O_w$ the active protected sessions, and $P_w$ the pending sessions that have been admitted but are not yet visible in the latest cache snapshot:
\begin{equation}
\textsc{Admit}(r,w)=
\begin{cases}
1, & s_r\in O_w\cup P_w,\\
1, & |O_w|+|P_w| < K_w(t),\\
0, & \text{otherwise}.
\end{cases}
\end{equation}
Requests whose sessions are already in $O_w$ or $P_w$ continue without consuming additional cap headroom.
A request whose session is in neither set must acquire cap headroom; otherwise, the router holds and periodically re-evaluates it.
Pending sessions prevent over-admission between snapshots, and held sessions remain in the demand ledger.
For grouped rollouts, the recheck delay decreases as more siblings from the same prompt group become visible to the scheduler.

For a continuation that must reacquire admission, let $L_r$ be its prompt length, $H_r$ its reusable prefix, $F$ the prefill throughput, $q_{\mathrm{cold}}$ the cold-route queueing delay, $q_{\mathrm{hit}}$ the locality-preserving queueing delay, and $\Delta_r$ the hold delay.
Holding for the cached instance is preferable to immediate cold routing whenever
\begin{equation}
\Delta_r + q_{\mathrm{hit}} - q_{\mathrm{cold}} < \frac{H_r}{F}.
\label{eq:hold-condition}
\end{equation}
At prefill saturation, request rate is bounded by $F/\mathbb{E}[L_r-H_r]$; admitting additional sessions cannot improve this bound if KV churn reduces $H_r$.

Let $\bar{k}_w(t)$ be a conservative reference length computed from a moving window of completed-session lengths and current observations.
If $C_w$ KV blocks are reserved for protected locality, the baseline cap is $K_w^{\mathrm{base}}(t)=\lfloor C_w/\bar{k}_w(t)\rfloor$.
The operational cap satisfies $K_w(t)\leq K_w^{\mathrm{base}}(t)$.
The baseline follows session length.
A confirmed decline in prefix-cache hit rate or a persistent waiting queue triggers overload pressure, which temporarily contracts the operational cap before gradual recovery.
Overload pressure is therefore a fast correction when the length reference lags, not the normal source of the cap.

Prefix-hit degradation is one overload signal and is evaluated relative to recent achievable behavior rather than by a fixed hit-rate threshold.
For instance $w$, the scheduler compares the observed hit rate $\hat{h}_w(t)$ with a reference $h^\star_w(t)$ learned from recent healthy windows:
\begin{equation}
G_w(t)=\max\{0,h^\star_w(t)-\hat{h}_w(t)\},
\end{equation}
Confirmation and slope checks filter transient changes before cap reduction or recovery.

\subsection{Workload-aware session caps}

A global session cap controls aggregate instance load but treats all sessions as interchangeable.
In a mixed pool, the same session count can represent different KV demand, and a dominant class can occupy most capacity available to new sessions.
MISA therefore adds class-specific soft quotas.
For inference instance $w$ and workload class $b$, $N_{w,b}$ is the instantaneous number of distinct unfinished session demands visible when evaluating $w$, rather than a time-averaged resident population.
Active, pending, held, and candidate sessions each contribute one count; held and candidate sessions do not consume the cap until admission, when the selected instance records them as pending sessions.
HOLD delays work without removing its demand from the ledger.
$\bar{k}_b$ estimates resident blocks separately for each workload class using conservative length priors, observed session blocks, and bounded class histories.

Ignoring residency duration for the moment, the spatial KV demand of class $b$ is
\begin{equation}
S_{w,b}=N_{w,b}\,\bar{k}_b.
\label{eq:spatial-demand}
\end{equation}
With protected KV budget $C_w$, MISA assigns a block quota proportional to this spatial demand and converts that quota back into a session cap:
\begin{equation}
M^{\mathrm{space}}_{w,b}=C_w\frac{S_{w,b}}{\sum_{b'}S_{w,b'}},
\qquad
K^{\mathrm{space}}_{w,b}=\max\!\left\{1,\left\lfloor\frac{M^{\mathrm{space}}_{w,b}}{\bar{k}_b}\right\rfloor\right\}.
\label{eq:workloadcap}
\end{equation}
Here, $K^{\mathrm{space}}_{w,b}$ gates new or rebound sessions.
These per-class caps partition the protected capacity represented by the global cap; a new session is admitted only when both the global cap and its class cap have headroom.
Bounded length histories update class quotas as recent workload lengths change.

\subsection{KV-cache residency-time weighting}

Equation~\eqref{eq:workloadcap} separates workload classes but estimates demand from KV footprint alone, treating equal block footprints as equal demand regardless of how long they remain resident.
Let $\widehat{T}_b$ denote the recent class-level estimate of the observed session duration over which useful KV state remains attributable to class $b$.
For single-request samples, this spans admission through completion; for a multi-turn agent session, it also includes tool-execution intervals between model turns.

MISA-T replaces spatial demand with the block-time demand proxy
\begin{equation}
R_{w,b}=S_{w,b}\widehat{T}_b=N_{w,b}\bar{k}_b\widehat{T}_b.
\label{eq:block-time-demand}
\end{equation}
If class $b$ receives block quota $M_{w,b}$, its estimated backlog drain time is
\begin{equation}
\tau_{w,b}\approx\frac{N_{w,b}\bar{k}_b\widehat{T}_b}{M_{w,b}}.
\label{eq:backlog-drain-time}
\end{equation}
Allocating quota in proportion to $R_{w,b}$ therefore approximately balances these estimated drain times across classes.
It then allocates the protected block budget and derives the class cap as
\begin{equation}
M^{\mathrm{time}}_{w,b}=C_w\frac{R_{w,b}}{\sum_{b'}R_{w,b'}},
\qquad
K^{\mathrm{time}}_{w,b}=\max\!\left\{1,\left\lfloor\frac{M^{\mathrm{time}}_{w,b}}{\bar{k}_b}\right\rfloor\right\}.
\label{eq:residency-quota}
\end{equation}
Here, $\widehat{T}_b$ affects only the budget share, while $\bar{k}_b$ converts allocated blocks into a session cap.
Because $R_{w,b}$ includes current outstanding demand, an accumulating class backlog receives a larger quota, while its share falls as that backlog is drained.

The controller estimates $\widehat{T}_b$ online from recently completed sessions, independently for each class.
A shared neutral prior stabilizes early quotas before each class has sufficient observations.
The product $\bar{k}_b\widehat{T}_b$ conservatively approximates the block-seconds integral $\int k_i(t)\,dt$.
MISA denotes the spatial allocation in Eq.~\eqref{eq:workloadcap}; \sys{} denotes the residency-weighted allocation in Eq.~\eqref{eq:residency-quota}.

\subsection{Session-level cache accounting}

The routing layer receives two complementary views of cache state from each inference instance.
A session snapshot records running and recently completed sessions, their estimated total and cached KV blocks, lifecycle timestamps, and the instance block capacity.
It is constructed from session-tagged request events and observed token lengths, and session admission uses it to identify continuations, estimate protected occupancy, and compute admission caps.

Prefix location is tracked separately through block hashes produced by the inference engine's prefix cache.
The serving layer reports hash additions and removals, from which the router maintains a mapping from each prefix hash to the instances that currently cache it.
For an incoming request, the longest matching prompt-block hash determines the reusable prefix on each instance; eviction or cache reset removes the corresponding hashes and invalidates that location.

\subsection{Implementation summary}

\sys{} is implemented as a routing-layer admission extension that can run in a separate scheduler or be colocated with a serving gateway.
It filters instances using runtime metrics, session snapshots, and workload metadata, after which the existing cache-aware placement policy selects among instances with admission headroom.

\section{Evaluation}
\label{sec:evaluation}

Our evaluation addresses three questions.
First, can adaptive session admission prevent prefix-cache hit-rate collapse under high rollout concurrency?
Second, do workload-aware caps and residency-time weighting improve serving efficiency beyond class-agnostic admission?
Third, do these serving improvements reduce end-to-end iteration time while preserving the trainer-defined mixture?
We use fixed-checkpoint rollout-only experiments to isolate the first two effects and matched 50-iteration Step3.7 training runs to evaluate the third.

All serving metrics are computed over fixed wall-clock windows.
RPM counts successful inference requests per minute, while rollout throughput counts complete rollout samples returned to the trainer per minute.
Prefix hit rate is the fraction of queried prompt tokens found in the prefix cache.
Prefill TPS denotes effective prompt-token throughput and includes prefix-hit tokens.
Because absolute request and token rates depend on model size, hardware, and serving implementation, the main text emphasizes relative changes from vLLM Router.
The end-to-end evaluation additionally compares iteration time and consumed workload mixture at the same training horizon.
We report end-to-end and rollout-only measurements separately because weight synchronization in end-to-end training periodically resets and re-warms the prefix cache.

\subsection{Experimental setup}

\paragraph{End-to-end RL training.}
We compare \sys{} with vLLM Router on Step3.7 for 50 iterations under the same model initialization, sampling stream, target mixture, and verifier/reward configuration.
Step3.7 is a 196B-A11B sparse-MoE model.
The deployment uses six NVIDIA H200 nodes for training and four for rollout inference; each H200 GPU has 141~GB of HBM3e memory, and each inference node runs one TP=8 replica.
In a preliminary unbounded-admission stress test, prefix hit rate fell below 5\%, repeated prefill dominated the queue, and both prefill and decode throughput decreased markedly.
For the matched end-to-end runs, both policies use the same fixed per-instance concurrency ceiling, selected before comparison to sustain high GPU KV utilization while retaining headroom for reusable prefixes.
This gives vLLM Router a competitive static operating point for comparison with adaptive admission.

\paragraph{Rollout-only ablation.}
The rollout-only benchmark generates online mixed rollouts at a fixed model checkpoint, isolating serving-side scheduling from weight updates and cache resets.
Each reported value is the arithmetic mean of three independent runs.
The admission-controlled variants use the same controller parameters across both model and hardware configurations.
Step3.7 uses two NVIDIA H200 inference nodes, each running one TP=8 replica; Qwen3.6-35B-A3B uses 16 NVIDIA H100 GPUs with 80~GB of HBM3 memory each, organized as four TP=4 replicas.
For the vLLM Router baseline, we perform a coarse static-concurrency sweep with increments of 32 for Step3.7 and 16 for Qwen3.6-35B-A3B; the best observed static concurrency is 128 and 48, respectively.
We additionally report a capacity-safe configuration based on the KV-cache capacity required by a 128K-token context.
The high-load Qwen3.6-35B-A3B diagnostic gives vLLM Router the same permissive concurrency ceiling used by Session Admission, MISA, and \sys{}; unlike these admission policies, vLLM Router does not dynamically limit admitted-session concurrency.
Each fanout sample has a stable session identity and workload class.
We compare:
\begin{enumerate}
    \item \textbf{vLLM Router:} the cache-aware policy favors prefix matches under balanced load and lower-load instances under imbalance, without admission control.
    \item \textbf{Session Admission:} adaptive class-agnostic admission that protects existing continuations and holds new sessions when the global cap is full.
    \item \textbf{MISA:} session admission with workload-specific caps derived from class demand.
    \item \textbf{\sys{}:} MISA with class-level KV-residency-time weighting.
\end{enumerate}
We report completed rollout samples per minute, request RPM, prefix hit rate, effective prefill TPS, and decode TPS.
Together, these variants isolate adaptive admission, workload-aware capacity allocation, and residency-time-aware KV accounting.

\subsection{End-to-end training results}

Table~\ref{tab:e2e-results} summarizes the matched Step3.7 end-to-end comparison.
All serving deltas are relative to vLLM Router.

\begin{table}[H]
    \centering
    \caption{End-to-end Step3.7 results over 50 iterations.}
    \label{tab:e2e-results}
    \tablefont
    \setlength{\tabcolsep}{1.4pt}
    \begin{tabular*}{\linewidth}{@{\extracolsep{\fill}}lrrrrrrrrr}
        \toprule
        & \multicolumn{5}{c}{Change vs. baseline (\%)} & & \multicolumn{3}{c}{50-iteration mix (\%)} \\
        \cmidrule(lr){2-6}\cmidrule(lr){8-10}
        Router & Iter. time & RPM & Sample rate & Prefill TPS & Decode TPS & Prefix hit & Agent & RLVR & RLHF \\
        \midrule
        \vllmroutercell & 0.0\% & 0.0\% & 0.0\% & 0.0\% & 0.0\% & 74.5\% & 48.3 & 29.4 & 22.4 \\
        \sys{} & \textbf{-22.8\%} & \textbf{+32.5\%} & \textbf{+35.6\%} & \textbf{+33.6\%} & \textbf{+38.0\%} & \textbf{96.2\%} & 45.9 & 33.3 & 20.8 \\
        \bottomrule
    \end{tabular*}
\end{table}

Under the same 50-iteration training horizon, \sys{} increases rollout throughput by $35.6\%$, reduces mean iteration time by $22.8\%$, and raises prefix-cache hit rate from $74.5\%$ to $96.2\%$.
Request, prefill, and decode throughput also improve, as summarized in Table~\ref{tab:e2e-results}.
These end-to-end rates include periodic weight synchronization: inference replicas reset prefix caches after loading updated weights, and subsequent requests perform cold prefill until reusable state accumulates again.

\begin{figure}[!htbp]
    \centering
    \includegraphics[width=0.98\linewidth]{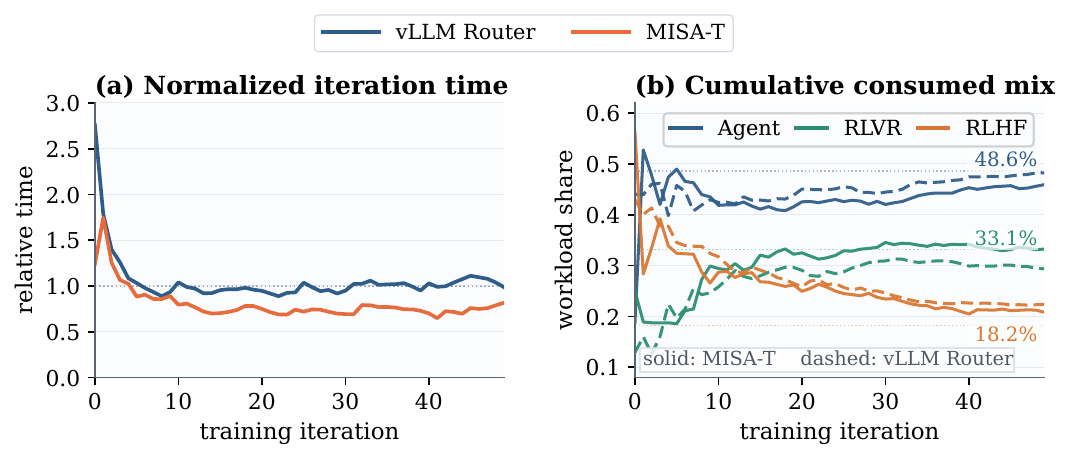}
    \caption{End-to-end progress over 50 iterations.}
    \label{fig:end-to-end-consistency}
\end{figure}

The target Agent/RLVR/RLHF mixture is $48.6/33.1/18.2\%$.
Over 50 iterations, \sys{} consumes $45.9/33.3/20.8\%$, with a total-variation distance of $2.71$ percentage points, compared with $48.3/29.4/22.4\%$ and $4.14$ points under vLLM Router.
The trainer remains responsible for specifying the target distribution; the routing policy only avoids introducing additional serving-side imbalance.
At the same training horizon, the absolute pass@4 difference between \sys{} and vLLM Router is below $0.5$ percentage points on SWE-Pro, SWE-Verified, and SWE-MTLG.
Figure~\ref{fig:end-to-end-consistency} reports smoothed iteration time normalized by the mean vLLM Router value and cumulative consumed mixture over the matched 50-iteration runs.
Solid and dashed curves denote \sys{} and vLLM Router, while dotted lines mark the trainer targets.

\raggedbottom
\subsection{Rollout-only ablation}

Table~\ref{tab:rollout-results} compares the four scheduling policies on Step3.7 and Qwen3.6-35B-A3B at fixed checkpoints.
All deltas are relative to the sweep-tuned vLLM Router on the same model.

\begin{table}[!htbp]
    \centering
    \caption{Rollout-only serving results.}
    \label{tab:rollout-results}
    \tablefont
    \setlength{\tabcolsep}{1.8pt}
    \begin{tabular*}{\linewidth}{@{\extracolsep{\fill}}llrrrrr}
        \toprule
        & & \multicolumn{4}{c}{Change vs. baseline (\%)} & \\
        \cmidrule(lr){3-6}
        Model & Router & Sample rate & RPM & Prefill TPS & Decode TPS & Prefix hit \\
        \midrule
        Step3.7 & \mbox{vLLM Router (128K-safe)} & -21.7\% & -27.8\% & -29.5\% & -22.1\% & 97.1\% \\
        Step3.7 & \mbox{vLLM Router (sweep)} & 0.0\% & 0.0\% & 0.0\% & 0.0\% & 95.9\% \\
        Step3.7 & Session Admission & +11.7\% & +22.7\% & +20.5\% & +16.8\% & 97.1\% \\
        Step3.7 & MISA & +23.3\% & +28.8\% & +25.4\% & +19.8\% & 97.3\% \\
        Step3.7 & \sys{} & \textbf{+53.3\%} & \textbf{+45.5\%} & \textbf{+40.4\%} & \textbf{+26.5\%} & \textbf{97.8\%} \\
        \midrule
        Qwen3.6-35B-A3B & \mbox{vLLM Router (128K-safe)} & -12.2\% & -10.2\% & -9.0\% & -6.8\% & 95.9\% \\
        Qwen3.6-35B-A3B & \mbox{vLLM Router (sweep)} & 0.0\% & 0.0\% & 0.0\% & 0.0\% & 92.4\% \\
        Qwen3.6-35B-A3B & Session Admission & +19.3\% & +23.7\% & +23.4\% & +7.8\% & \textbf{96.3\%} \\
        Qwen3.6-35B-A3B & MISA & +30.9\% & +28.9\% & +30.2\% & +12.3\% & 79.2\% \\
        Qwen3.6-35B-A3B & \sys{} & \textbf{+43.6\%} & \textbf{+37.5\%} & \textbf{+36.5\%} & \textbf{+18.3\%} & 95.3\% \\
        Qwen3.6-35B-A3B & \vllmrouterhighcell & -55.2\% & -52.5\% & -57.3\% & -38.5\% & 4.5\% \\
        \bottomrule
    \end{tabular*}
\end{table}

Relative to sweep-tuned vLLM Router, \sys{} improves rollout throughput by $53.3\%$ on Step3.7 and $43.6\%$ on Qwen3.6-35B-A3B, with prefix-cache hit rates of $97.8\%$ and $95.3\%$, respectively.
Request throughput also increases by $45.5\%$ and $37.5\%$, respectively.
Session Admission, MISA, and \sys{} dynamically regulate admitted-session concurrency, so they use a common permissive concurrency ceiling.
In the additional high-load Qwen3.6-35B-A3B configuration, vLLM Router uses the same ceiling but reaches only a $4.5\%$ hit rate.
This reproduces the prefill-eviction feedback loop described in Section~2.5; every admission-based policy interrupts the loop and improves both request throughput and prefix reuse.
\sys{} provides the largest rollout-throughput gain on both models.

The ablation follows the role of the three controls introduced in Section~4.
vLLM Router improves placement but admits new sessions without accounting for their aggregate KV demand.
Session Admission bounds that demand and protects continuations, but uses one class-agnostic pool.
MISA separates workload classes using demand and KV footprint, while \sys{} additionally accounts for observed residency duration.
Relative to Session Admission, MISA improves rollout throughput by $10.4\%$ on Step3.7 and $9.7\%$ on Qwen3.6-35B-A3B.
Adding residency-time weighting improves it by a further $24.3\%$ and $9.7\%$, respectively.

\subsection{Controller behavior}

\begin{figure}[H]
    \centering
    \includegraphics[width=0.90\linewidth]{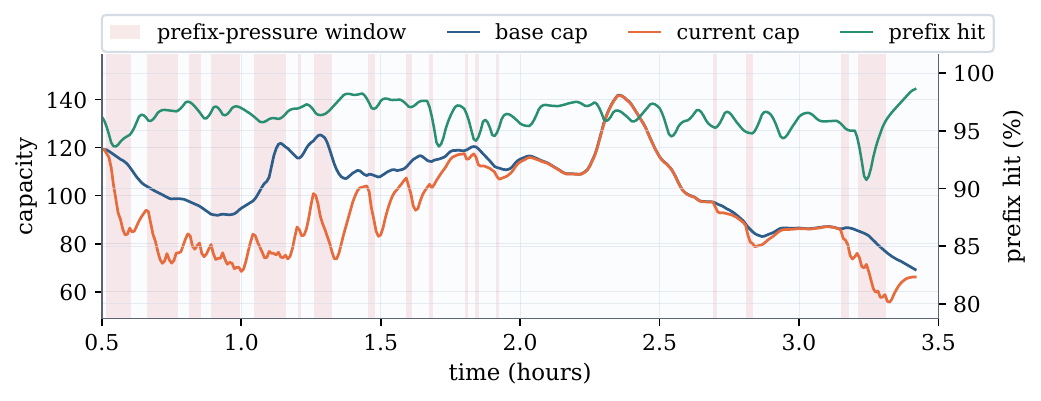}
    \caption{\sys{} admission control on Qwen3.6-35B-A3B.}
    \label{fig:scheduler-pressure}
\end{figure}

Figure~\ref{fig:scheduler-pressure} shows admission-cap adaptation under prefix-cache pressure.

The base cap follows the moving session-length reference, while the current cap gates new-session admission.
When confirmed prefix-hit degradation indicates that this reference is lagging current demand, the controller temporarily lowers the current cap; red bands mark these pressure windows relative to recent healthy hit rates.
Long-tailed session growth produces more corrections early in the run, but the caps align and pressure becomes less frequent as completed-session observations update the reference.

\subsection{Compatibility with CPU KV offloading}

We additionally run a Step3.7 rollout-only experiment with CPU KV-cache offloading enabled.

\begin{figure}[!t]
    \centering
    \includegraphics[width=0.99\linewidth]{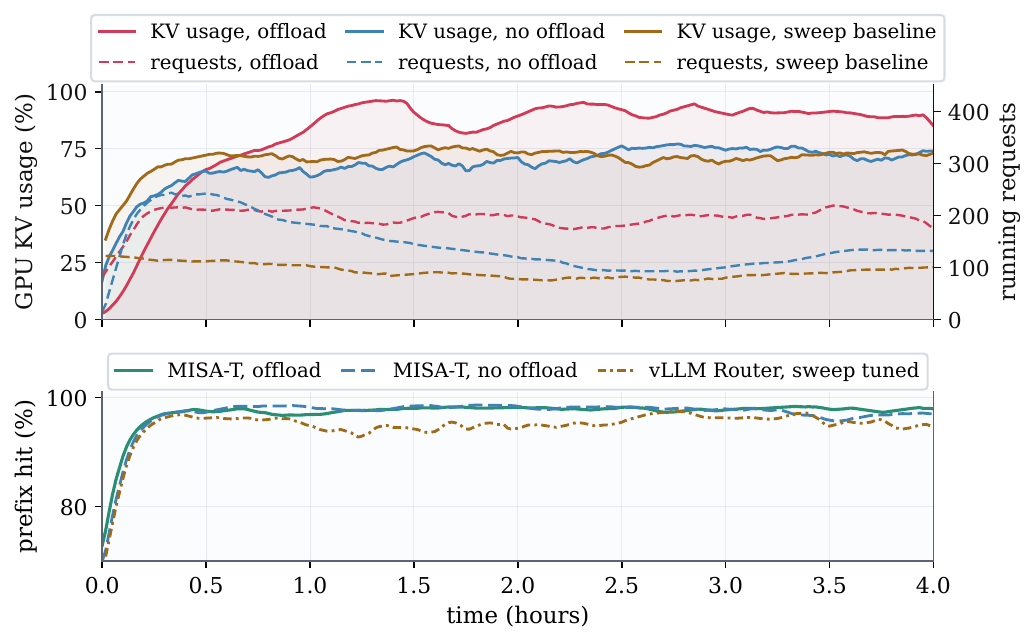}
    \caption{\sys{} with CPU KV-cache offloading on Step3.7.}
    \label{fig:cpu-offload-controller}
\end{figure}

The offload and no-offload runs use the same MISA-T concurrency configuration; CPU KV-cache offloading is the only changed serving option.
The CPU tier is configured with 1.3~TB of KV-cache capacity and maintains a backup copy of all KV-cache state resident on the GPUs.
\sys{} continues to control logical session admission, while the offloading layer manages the physical cache copies without requiring changes to the routing policy.
This experiment tests whether routing-layer admission remains effective when reusable KV state is backed by a CPU tier, while allowing the active GPU KV cache to operate close to capacity.

After the initial warm-up, Figure~\ref{fig:cpu-offload-controller} shows that GPU KV-cache usage remains close to or above $90\%$ through most of the steady-state interval.
The corresponding no-offload trace remains lower, demonstrating that CPU KV cache offloading is compatible with \sys{}, allows the GPU KV cache to remain highly utilized, and improves mean per-replica RPM by $35.6\%$.

\section{Discussion}

\paragraph{Adaptive admission versus static concurrency.}
\sys{} retains the outer rollout-concurrency limit, which bounds total work in flight, and adaptively controls how many cache-resident sessions enter each inference instance.
It does not seek a globally optimal outer limit.
Instead, it reduces reliance on repeated static-concurrency searches as sequence lengths, residency patterns, and serving demand vary across training algorithms, workload configurations, and model checkpoints.

\paragraph{Compatibility with KV offloading.}
KV-offloading systems extend effective cache capacity by moving selected KV state outside GPU memory~\citep{lee2024infinigen,sun2025shadowkv}.
\sys{} and KV-cache offloading act at complementary layers: the router controls logical session admission, while the offloading layer determines whether reusable KV state resides on GPU or CPU memory.
The experiment in Section~5.5 shows that CPU-backed reuse can increase active GPU KV occupancy while admission control retains prefix locality.

\section{Conclusion}

Mixed RL rollout serving requires more than prefix-aware instance selection.
Under high concurrency, the routing layer must control how many sessions compete for KV-cache capacity and allocate protected capacity across workloads with different KV footprints and residency times, while leaving the target workload mixture under trainer control.
\sys{} addresses these requirements through adaptive session admission, workload-aware session caps, and residency-time-aware KV accounting.

Rollout-only ablations show that \sys{} improves rollout throughput over sweep-tuned vLLM Router by $53.3\%$ on Step3.7 and $43.6\%$ on Qwen3.6-35B-A3B while retaining high prefix-cache hit rates.
In the matched 50-iteration Step3.7 run, \sys{} increases rollout throughput by $35.6\%$, reduces mean iteration time by $22.8\%$, and decreases mixture deviation from $4.14$ to $2.71$ percentage points.
Together, these results show that routing-layer admission control can improve mixed-rollout serving efficiency while maintaining comparable task scores.

\paragraph{Limitations.}
\sys{} assumes that requests carry workload labels and relies on timely serving-state reports from inference instances.
Delayed or incomplete snapshots can temporarily reduce the accuracy of KV-demand estimates and admission caps.

\bibliographystyle{plainnat}

\end{document}